\documentclass[aps,nofootinbib,superscriptaddress,floatfix,showpacs,showkeys,onecolumn,10pt]{revtex4-1}
\usepackage{amssymb}
\usepackage{amsmath, bm, graphicx,color,epstopdf,graphics}
\usepackage{subfigure}
\usepackage{amsmath}
\usepackage{amsfonts}
\usepackage{graphicx}

\newcommand{\be}{\begin{eqnarray}}
\newcommand{\ee}{\end{eqnarray}}

\renewcommand{\d}{\mbox{${\rm d}$}}

\newcommand{\gn}{G_{\rm N}}
\newcommand{\rh}{r_{\rm H}}

\newcommand{\Rh}{R_{\rm H}}

\begin{document}
\title{The marginally trapped surfaces in spheroidal spacetimes}
\author{Rehana Rahim}
\email{E-mail: rehana.rahim8@gmail.com}
\affiliation{Dipartimento di Fisica e Astronomia, Universit\`a di Bologna, via Irnerio~46, 40126 Bologna, Italy}
\affiliation{Department of Mathematics, Quaid-i-Azam University, Islamabad, Pakistan}
\author{Andrea Giusti}
\email{E-mail: agiusti@bo.infn.it}
\affiliation{Dipartimento di Fisica e Astronomia, Universit\`a di Bologna, via Irnerio~46, 40126 Bologna, Italy}
\affiliation{I.N.F.N., Sezione di Bologna, IS - FLAG, via B.~Pichat~6/2, 40127 Bologna, Italy}
\affiliation{Arnold Sommerfeld Center, Ludwig-Maximilians-Universit\"at, Theresienstra\ss e~37, 80333~M\"unchen, Germany}
\author{Roberto Casadio}
\email{E-mail: casadio@bo.infn.it}
\affiliation{Dipartimento di Fisica e Astronomia, Universit\`a di Bologna, via Irnerio~46, 40126 Bologna, Italy}
\affiliation{I.N.F.N., Sezione di Bologna, IS - FLAG, via B.~Pichat~6/2, 40127 Bologna, Italy}
\begin{abstract}
We study the location of marginally trapped surfaces in space-times resulting from an axial deformation
of static isotropic systems, and show that the Misner-Sharp mass evaluated on the corresponding
undeformed spherically symmetric space provides the correct gravitational radius to locate
the spheroidal horizon.
\end{abstract}


\maketitle

\section{Introduction and motivation}
\label{intro}
According to General Relativity, black holes are portions of Lorentzian manifolds characterised
by the existence of an event horizon, from within which no signals can ever escape.
In more general gravitating systems, the local counterpart of the event horizon is given by
a marginally outer trapped surface (MOTS)~\cite{Wald,senovilla}, which can be naively understood as the location
where the escape velocity equals the speed of light at a given instant. 
If the system approaches an asymptotically static regime, the outermost MOTS
should then become the future event horizon, like it happens in the very simple Oppenheimer-Snyder
model~\cite{OS}.
\par
More formally, a MOTS occurs where the expansion of outgoing null geodesics
vanishes~\cite{ashtekar,HE,Wald,Hayward:1993ma,senovilla}.
In general, the expansion scalars associated with outgoing and ingoing geodesics
are respectively given by
\begin{equation}
\Theta_{\bm{\ell}}
=
q^{\mu\nu}\,\nabla_\mu l_{\nu}
\ ,
\qquad
\Theta_{\bm{n}}
=q
^{\mu\nu}\,\nabla_\mu n_{\nu}
\ ,
\label{h10}
\end{equation}
where $\mu,\nu=0,\ldots,3$ and
\be
q_{\mu \nu}=g_{\mu \nu}+l_{\mu}\,n_{\nu}+n_{\mu}\,l_{\nu}
\label{2-g}
\ee
is the metric induced by the space-time metric $g_{\mu\nu}$ on the 2-dimensional
space-like surface formed by spatial foliations of the null hypersurface generated by
the outgoing null tangent vector $\bm{\ell}$ and the ingoing null tangent vector $\bm{n}$.
This 2-dimensional metric is purely spatial and has the following properties
\begin{equation}
q_{\mu\nu}\,\ell^{\mu}
=
q_{\mu\nu}\,\ell^{\nu}=0
\ ,
\quad
q^\mu_{\ \mu}=2
\ ,
\quad
q^\mu_{\ \lambda}\,q^\lambda_{\ \nu}
=
q^\mu_{\ \nu}
\ ,
\label{h11}
\end{equation}
where $q^\mu_{\ \nu}$ represents the projection operator onto the 2-space orthogonal
to $\bm{\ell}$ and $\bm{n}$.
In particular, one finds~\cite{Wald}
\begin{equation}
\Theta_{\bm{\ell}}
=
l^{\mu}\,\frac{\partial_\mu \sqrt{q}}{\sqrt{q}}
\ ,
\qquad
\Theta_{\bm{n}}
=
n^{\mu}\,\frac{\partial_\mu \sqrt{q}}{\sqrt{q}}
\ ,
\label{ThetaA}
\end{equation}
where $q$ is the determinant of the 2-dimensional metric $q_{\mu\nu}$ on surfaces orthogonal
to the null congruences. 
This makes it apparent that the expansion scalars describe how the transverse area spanned
by congruences changes along their evolution.
\par
Given these definitions, it is clear that the study of marginally trapped surfaces in any realistic system
is a very complex topic, and determining their existence and location is in general possible
only by means of numerical methods. 
However, for the particular case of a spherically symmetric self-gravitating source,
one can employ the gravitational radius, and the equivalent Misner-Sharp
mass function.
We recall that we can always write a spherically symmetric line element as
\be
\d s^2
=
g_{ij}(x^k)\,\d x^i\,\d x^j
+
r^2(x^i)\left(\d\theta^2+\sin^2\theta\,\d\phi^2\right)
\ ,
\label{metric}
\ee
where $x^i=(x^1,x^2)$ parametrise surfaces of constant angular coordinates
$\theta$ and $\phi$.
For the metric~\eqref{metric}, the gradient $\nabla_i r$ is orthogonal to surfaces
of constant area $\mathcal{A}=4\,\pi\,r^2$, and one finds that the product~\cite{senovilla,dafermos}
\be
\Theta_{\bm{\ell}} \,\Theta_{\bm{n}}
\propto
g^{ij}\,\nabla_i r\,\nabla_j r
\label{exp0}
\ee
precisely vanishes on marginally trapped surfaces.
Moreover, if we set $x^1=t$ and $x^2=r$, and denote the matter density as
$\rho=\rho(t,r)$, Einstein's field equations yield the solution
\be
g^{rr}=1-\frac{\rh(t,r)}{r}
\ ,
\label{grr}
\ee
where~\footnote{We shall use units with $\gn=c=1$.}
\be
\rh(t,r)
=
2\,{m(t,r)}
\label{hoop}
\ee
is the gravitational radius determined by the Misner-Sharp mass function \cite{Hayward:1994bu}
\be
m(t,r)
=
4\,\pi
\int_0^r \rho(t,\bar r)\,\bar r^2\,\d \bar r
\ .
\label{M}
\ee
According to Eq.~\eqref{exp0}, a MOTS then exists where $g^{rr}=0$,
or where the gravitational radius satisfies 
\be
\rh(t,r)= r
\ ,
\label{Ehor}
\ee
for $r>0$.
If the source is surrounded by the vacuum, the Misner-Sharp mass asymptotically approaches the
Arnowitt-Deser-Misner (ADM) mass of the source, $m(t,r\to\infty)=M$,
and the gravitational radius likewise becomes the Schwarzschild radius $\Rh= 2\,M$.
To summarise, the relevant properties of the Misner-Sharp mass~\eqref{M} are that 
{\em i)\/} it only depends on the source energy density and
{\em ii)\/} it allows one to locate the (time-dependent) MOTS via Eq.~\eqref{hoop}.
\par
In quantum physics, the energy density that defines the Misner-Sharp mass $m$
(and ADM mass $M$) becomes a quantum observable and one expects the gravitational radius
to admit a similar description. 
The horizon quantum mechanics (HQM) was in fact proposed~\cite{fuzzyh}
in order to describe the ``fuzzy'' Schwarzschild (or gravitational) radius of a localised quantum
source, by essentially lifting Eq.~\eqref{hoop} to a quantum constraint acting on the state
vectors of matter and the gravitational radius.
In this respect, the HQM differs from most other attempts in which the gravitational degrees
of freedom of the horizon, or of the black hole metric, are instead quantised independently of the state
of the source.
It however follows that, in order to extend the HQM to non-spherical systems, we need to identify a mass
function from which the location of a MOTS, $r=\rh$, can be uniquely determined
and which depends only on the state of the matter source, like the Misner-Sharp
mass~\eqref{M} for isotropic sources.
The latter property is crucial in a perspective in which one would eventually like to recover
the geometric properties of space-times from the quantum state of the whole matter-gravity system.
\par
Since we are interested in generalising the above quantum description to non-spherical
sources, in this work we shall first try and generalise the classical analysis of marginally trapped surfaces
to systems with a slightly spheroidal symmetry. 
Moreover, since it is hardly possible to describe analytically such systems if they evolve in time,
we shall consider static configurations as simple case studies.
In particular, we shall deform a static and spherically symmetric space-time, and study the 
location of marginally trapped surfaces perturbatively in the deformation parameter. 
In this respect, it is worth stressing that the assumption of staticity will ultimately lead to matter
distributions which break some of the energy conditions.
The cases presented here are therefore only intended to serve as toy models,
whose purpose is to shed some light on the possible relation between these small perturbations
and a mass function.
Consequently, the development of a more precise analysis for dynamical horizons is left
for future studies.
\par
Explicit expressions will be given for the deformed de~Sitter space-time.
We shall also consider the case of a spheroidal space-time which contains 
a source whose energy and pressure depart from such a symmetry.
In both cases, we will see that the location of marginally trapped surfaces is given by surfaces
of symmetry, and can therefore be determined by computing the Misner-Sharp mass
on the reference unperturbed (spherically symmetric) space-time.
The results of this analysis will serve in order to establish the adapted 
quantization rules for the HQM of such systems, but the whole quantum 
extension will be described in other publications (for some preliminary results,
see Ref.~\cite{letter}).
\section{Static spheroidal sources}
\label{Sec:spheroidal}
In this section, we will investigate how the particular description for static spherically symmetric 
systems extends to the case in which the symmetry is associated with (slightly) spheroidal surfaces.
We start from the spherically symmetric metric~\eqref{metric}, with $r$ the areal radius
constant on the 2-spheres of symmetry, and assume the time-dependence is negligible.
Einstein equations then yield the solution~\eqref{grr}, in which the now time-independent 
Misner-Sharp mass $m=m(r)$ is determined by a static density $\rho=\rho(r)$
according to Eq.~\eqref{M}.
We will always assume that the matter source also contains a (isotropic) pressure term, such that
the Tolman-Oppenheimer-Volkov equation of hydrostatic equilibrium is satisfied~\cite{stephani}.
We then change to (prolate or oblate) spheroidal coordinates and consider a localised source
of spheroidal radius $r=r_0$, say with mass $M_0$, surrounded by a fluid with the energy
density $\rho=\rho(r)$.
\begin{figure}[t!]	
\centering
\includegraphics[scale=0.3]{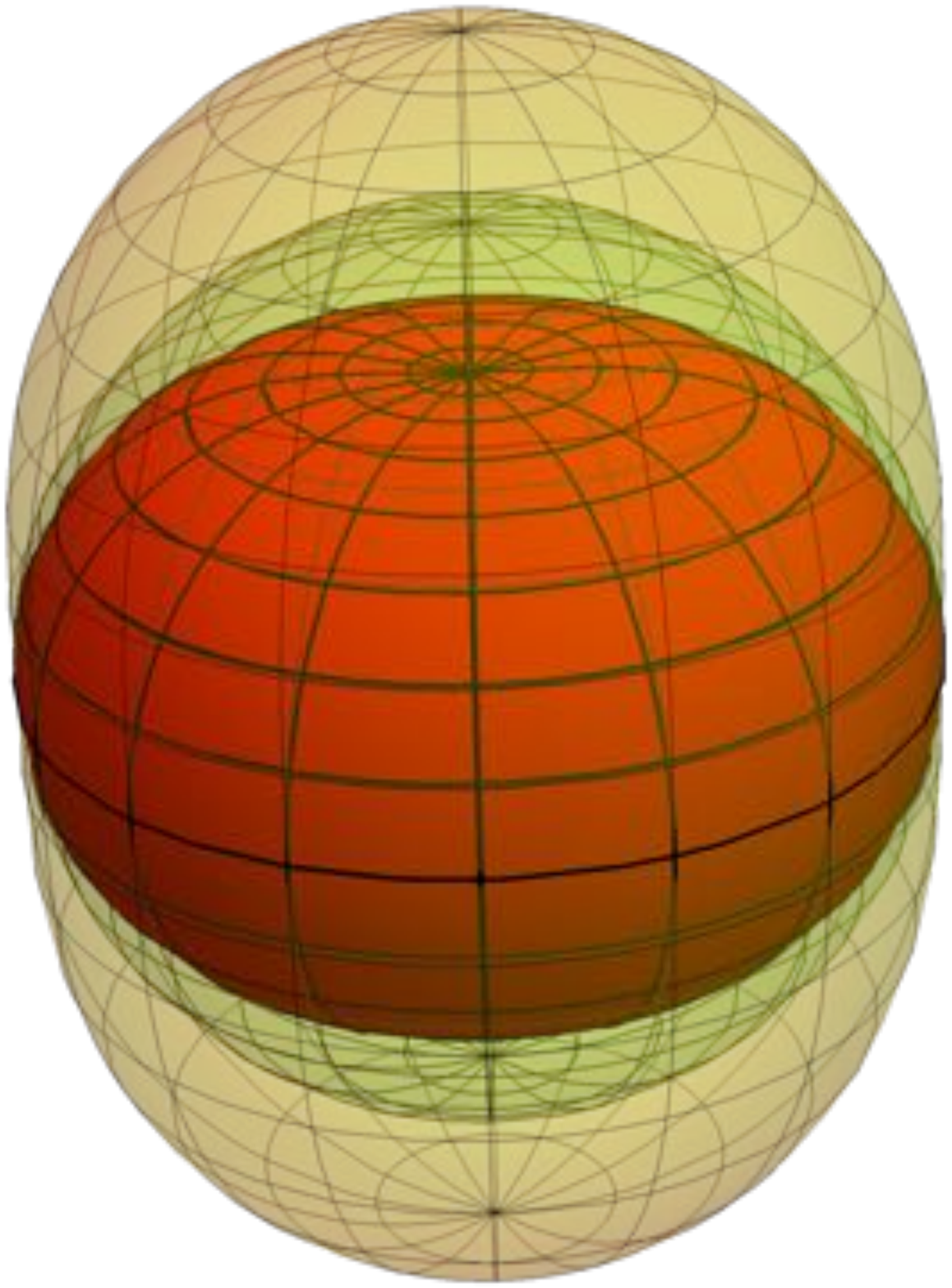}
\caption{Spheroids: prolate spheroid with $a^2>0$ (in yellow) compared to oblate spheroid with $a^2<0$
(in red) and to the reference sphere $a^2=0$ (in green).}
\label{f1}
\end{figure}
\par
The central source only serves the purpose to avoid discussing coordinate singularities at $r=0$.
In the interesting portion of space $r>r_0$, we assume the metric $g_{\mu\nu}$ is of the form
\be
\d s^{2}
&=&
-h(r,\theta;a)\, \d t^{2}
+\frac{1}{h(r,\theta;a)}
\left(\frac{r^{2}+a^{2}\cos{}^{2}\theta}{r^{2}+a^{2}}\right) \d r^{2}
\nonumber
\\
&&
+\left(r^{2}+a^{2}\cos{}^{2}\theta\right) \d\theta^{2}
+\left(r^{2}+a^{2}\right)\sin^{2}\theta\,\d\phi^{2}
\ ,
\qquad
\label{1}
\ee
where $h=h(r,\theta;a)$ is a function to be determined.
Surfaces of constant $r$ now represent ellipsoids of revolution, or spheroids, on which the
density is constant.
For $a^2>0$, the above metric can describe the space-time outside a prolate spheroidal source,
which extends more along the axis of symmetry than on the equatorial plane (see yellow surface
in Fig.~\ref{f1}).
In order to describe an oblate source, which is flatter along the axis of symmetry
(see red surface in Fig.~\ref{f1}),
we can simply consider the mapping $a\to i\,a$ (so that $a^2\to -a^2$).
It is also important to remark that a space-time equipped with the metric~\eqref{1} admits two trivial
Killing vectors, namely $\bm{\partial _t}$ and $\bm{\partial _\phi}$.
Furthermore, it is also easy to see that the vanishing of $g_{00}=-h(r,\theta;a)$ determines the location
of the Killing horizons for spacetimes belonging to this class~\footnote{In a more general, time-dependent
space-time, no such Killing structure would of course exist.}.
\par
For consistency, the energy-momentum tensor $T_{\mu\nu}$ of the source can be inferred from the
Einstein equations,
\begin{equation}
G_{\mu\nu}
=
R_{\mu\nu}
-\frac{1}{2}\,R\,g_{\mu\nu}
=8\,\pi\, T_{\mu\nu}
\ ,
\label{2}
\end{equation}
where $G_{\mu\nu}$ is the Einstein tensor, $R_{\mu\nu}$ the Ricci tensor and
$R$ the Ricci scalar.
However, we are only interested in ensuring that the energy density is spheroidally symmetric,
that is
\be
G^0_{\ 0}
=
-8\,\pi\,\rho(r)
\ ,
\label{eG00}
\ee
and we will therefore assume the necessary pressure terms are present in order to 
maintain equilibrium.
In order to solve Eq.~\eqref{eG00}, we change variable from the azimuthal angle $\theta$
to $x\equiv\cos\theta$, after which the line element reads
\be
\d s^{2}
&=&
-h(r,x;a)\,\d t^{2}
+\frac{1}{h(r,x;a)}
\left(\frac{r^{2}+a^{2}\,x^2}{r^{2}+a^{2}}\right) \d r^{2}
\nonumber
\\
&&
+\frac{r^{2}+a^{2}\,x^2}{1-x^2}\, \d x^2
+\left(r^{2}+a^{2}\right)(1-x^2)\,\d\phi^{2}
\ .
\label{gx}
\ee
Given the symmetry of the system, we can restrict the analysis to the upper half spatial volume 
$1\ge x\ge 0$ corresponding to $0\le\theta\le \pi/2$.
We then find
\begin{eqnarray}
G^0_{\ 0}
&=&
\frac{1}{4\,(r^{2}+a^{2}x^{2})^{3}\,h^2}
\Big\{4
\left[r^{4}+a^{2}r^{2}(4\,x^{2}-1)
+a^{4}\,x^{2}\,(1+x^{2})
\right]
h^3
\nonumber
\\
&&
+3\,(1-x^{2})(r^{2}+a^{2}x^{2})^{2}
\left({\partial_x h}\right)^{2}
\nonumber
\\
&&
+2\,(r^{2}+a^{2}x^{2})\,h
\left[x
\left(2\,r^{2}+a^{2}\left\{3\,x^{2}-1\right\}\right)\partial_x h
+(x^{2}-1)\,(r^{2}+a^{2}\,x^{2})\,\partial_x^{2}h\right]
\nonumber
\\
&&
+2\,h^{2}
\big[2
\left(a^{2}r^{2}\left\{1-4\,x^{2}\right\}
-a^{4}\,x^2\left\{1+x^{2}\right\}
-r^{4}
\right)
\nonumber
\\
&&
+r\left(r^{2}+a^{2}x^{2}\right)
\left(2\,r^{2}+a^{2}\left\{1+x^{2}\right\}
\right)
\partial_r h
\big]
\Big\}
\ ,
\label{G00}
\end{eqnarray}
so that Eq.~\eqref{eG00} appears to be a rather convoluted differential equation for
the unknown metric fuction $h=h(r,x;a)$.
\par
For completeness, we also show the remaining (non-vanishing) components of the
Einstein tensor, namely
\begin{equation}
G^1_{\ 1}
=
\frac{r^2 \, (h - 1)}{(r^2 + a^2 \, x^2)^2}
\ ,
\label{G11}
\end{equation}
\begin{equation}
G^2_{\ 2}
=
\frac{x \left[\left(a^2 \, x^2 + r^2\right) \partial _x h + 2\,a^2 \, x \, (h -1) \, h \right] + 
r \, \left(a^2 \, x^2 + r^2 \right) \, h \, \partial _r h}{2 \left(a^2 \, x^2 + r^2 \right)^2 \, h}
\label{G22}
\end{equation}
and
\begin{eqnarray}
G^3_{\ 3}
&=&
\frac{1}{4 \left(a^2 x^2+r^2\right)^3 h^2} 
\left\{
4 \, a^2\,h^3
\left[a^{2} x^{2} + r^2 \, (2\,x^{2}-1)\right]
\right.
\nonumber
\\
&&
+3\,(1-x^{2})(r^{2}+a^{2}x^{2})^{2}
\left({\partial_x h}\right)^{2}
\nonumber
\\
&&
+2\,(r^{2}+a^{2}x^{2})\,h
\left[x
\left(r^{2}+a^{2}\left\{2\,x^{2}-1\right\}\right)\partial_x h
+(x^{2}-1)\,(r^{2}+a^{2}\,x^{2})\,\partial_x^{2}h\right]
\nonumber
\\
&&
\left.
+2\,h^{2}
\left[
2\, a^2 \left(r^2-x^2 \left\{a^2+2 r^2\right\}\right)
+r \left(a^2+r^2\right) \left(a^2\, x^2+r^2\right) \partial_r h
\right]
\right\}
\ ,
\label{G33}
\end{eqnarray}
from which it is easy to obtain the complete energy-momentum tensor
for a generic metric of the form~\eqref{1} from the Einstein equations~\eqref{2}.
\par
We proceed by considering small departures from spherical symmetry, parameterised
by $a^2\ll r_0^2$, and expand all expressions up to order $a^2$.
In particular, the energy density must have the form
\be
\rho
\simeq
\rho_{(0)}(r)
+
a^2\,\rho_{(2)}(r)
\ ,
\ee
whereas the unknown metric function
\be
h
&\simeq&
h_{(00)}(r)
+
a^2
\left[h_{(20)}(r)+x^2\,h_{(22)}(r)\right]
\ ,
\nonumber
\\
&\simeq&
1-\frac{2\,m_{(00)}(r)}{r}
-
2\,a^2
\frac{m_{(20)}(r)+x^2\,m_{(22)}(r)}{r}
\ ,
\quad
\label{h}
\ee
where we introduced a Misner-Sharp mass function $m_{(00)}$, like in Eq.~\eqref{M},
for the zero order term and corrective terms $m_{(2i)}$ at order $a^2$ (with $i = 0, 2$
representing the polynomial order in $x$).
In fact, at zero order in $a$, Eq.~\eqref{eG00} reads
\be
{G_{(0)}}^0_{\ 0}
=
-\frac{2\,m_{(00)}'(r)}{r^2}
=
-8\,\pi\,\rho_{(0)}(r)
\ ,
\label{m00}
\ee
with primes denoting derivatives with respect to $r$.
The solution $m_{(00)}$ is correctly given by the relation~\eqref{M}.
\par
At first order in $a^2$, the component of the Einstein tensor in Eq.~\eqref{G00} contains two terms, 
\be
{G_{(2)}}^0_{\ 0}(r,x)
=
F(r)
+x^2\,L(r)
\ ,
\ee
where $F (r)$ and $L(r)$ do not dependent of $x$.
Since $\rho$ does not depend on $x$ by construction, we must have $L (r) = 0$, which yields
\be
m_{(22)}'
+
\left(1-\frac{2\,m_{(00)}}{r}\right)^{-1}\frac{3\,m_{(22)}}{r}
-
\frac{3}{2\,r^2}
\left(m_{(00)}'-\frac{5\,m_{(00)}}{3\,r}\right)
=
0
\ .
\label{m22}
\ee
Finally, we are left with
\be
F(r)
&=&
-\frac{2\,m_{(20)}'}{r^2}
-
\frac{m_{(00)}'}{r^4}
+
\frac{3\,m_{(00)}}{r^5}
+
\frac{2\,m_{(22)}}{r^3}
\left(1-\frac{2\,m_{(00)}}{r}\right)^{-1}
\nonumber
\\
&=&
-8\,\pi\,\rho_{(2)}
\ ,
\label{m20}
\ee
in which $m_{(00)}$ is determined by Eq.~\eqref{m00} and $m_{(22)}$ by Eq.~\eqref{m22}, respectively.
Eq.~\eqref{m20} can then be used to determine $m_{(20)}$.
\par
Once the metric function $h=h(r,x;a)$ is obtained, one can determine the locations of marginally trapped surfaces 
from the expansions of null geodesics defined in Eq.~\eqref{h10}.
It will then be interesting to compare the result with the solutions of the generalised Eq.~\eqref{hoop},
namely
\be
2\,m(\rh,x;a)=\rh(x)
\ ,
\label{MSrh}
\ee
where
\be
m(r,x;a)
\simeq
m_{(00)}(r)
+
a^2\left[m_{(20)}(r)+x^2\,m_{(22)}(r)\right]
\ ,
\qquad
\ee
is now the extended Misner-Sharp mass.
We also note that Eq.~\eqref{MSrh} is equivalent to
\be
h(\rh,x;a)
=
0
\ ,
\label{h=0}
\ee
which will be checked below with a specific example.
\par
We can just anticipate that we expect the location of the MOTS will respect the
spheroidal symmetry of the system and be thus given by the spheroidal deformation
of the isotropic horizon obtained in the limit for $a\to 0$.
From Eq.~\eqref{ThetaA}, since the metric $q_{\mu\nu}$ is static, this will happen if we
can show that the tangent to outgoing null geodesics
\be
\bm{\ell}
=
\ell^t\,\bm{\partial}_t
+
\ell^r\,\bm{\partial}_r
\sim
h\,\bm{\partial} _r
\ee
when acting on functions of $r$ alone, and $q$ is regular where $h=0$.
\section{Slightly spheroidal de~Sitter}
\label{s:desitter}
In order to proceed and find more explicit results, we shall now apply the above general
construction to the specific example of the spheroidally deformed de~Sitter metric.
\par
Like in the previous general treatment, we start by assuming the presence of an inner core
of radius $r=r_0$ and mass $M_0$, which is here surrounded by a fluid with energy density
\be
\rho(r)
=
\rho_{(0)}(r)
=
\frac{\alpha ^2}{4\,\pi\,r}
\ ,
\label{rhoEx1}
\ee
where $r>r_0$, and $\alpha$ is a positive constant independent of $a$ (so that $\rho_{(2)}=0$).
From Eq.~\eqref{m00}, we obtain
\be
m_{(00)}
=
M_0
+
\frac{\alpha^2 \, (r^2-r_0^2)}{2}
\ ,
\ee
which of course holds for $r>r_0$.
We further set $\alpha^2\,r_0^2\simeq M_0$, so that
\be
m_{(00)} (r)
\simeq
\frac{\alpha^2 \, r^2}{2}
\ .
\ee
This case admits a MOTS when $2 \, m_{(00)} (r) = r$, that is 
\be
r_{\rm H} = \alpha^{-2}
\ ,
\label{rha0}
\ee
which is just the usual horizon for the isotropic de~Sitter space.
\par
Next, Eq.~\eqref{m22} reads
\be
m_{(22)}'
+\frac{3\,m_{(22)}}{(1-\alpha ^2 \, r)\,r}
-\frac{\alpha^2}{4\,r}
=
0
\ ,
\ee
and admits the general solution
\be
m_{(22)} (r)
=
\frac{1 - \alpha^2 \, r}{8\,\alpha^4\,r^3}
\left\{1
-
(1 - \alpha^2 \, r)
\right.
&&
\!\!\!\!
\left[4 - 8\,\alpha^4\,(1 - \alpha^2 \, r)\,C_{(22)}
\right.
\nonumber
\\
&&
\!\!\!
\left.
\left.
+ 2 \, (1 - \alpha^2 \, r) \, \ln (1 - \alpha^2 \, r)
\right]
\right\}
\ ,
\ee
with $C_{(22)}$ an integration constant.
For $r \simeq \alpha^{-2}$, the general solution reduces to
\be
m_{(22)} (r)
=
\frac{\alpha^2}{8} \, \left( 1 - \alpha^2 \, r \right) + o [(1 - \alpha^2 \, r)^2]
\ .
\ee
We can then determine $m_{(20)}$ from Eq.~\eqref{m20}, which,
on employing the above expansion for $m_{(22)}$, reads
\be
m_{(20)}'
\simeq
\frac{3 \, \alpha^2}{8 \, r}
\ ,
\ee
and yields
\be
m_{(20)} (r)
\simeq
C_{(20)} + \frac{3 \, \alpha^2}{8} \, \ln( \alpha^2 r)
\ ,
\ee
with $C_{(20)}$ another integration constant.
\par
We then set $C_{(20)}=C_{(22)}=0$ for simplicity, and obtain
\be
m(r,x;a)
&\simeq&
\frac{\alpha^2 \, r^2}{2}
+ \frac{a^2 \, \alpha^2}{8} \, \left[ (1 - \alpha^2 \, r) \, x^2 + 3 \, \ln( \alpha^2 r) \right]
\ ,
\ee
for $r\simeq \alpha^{-2}$.
After substituting $m_{(00)}$, $m_{(20)}$ and $m_{(22)}$ into Eq.~\eqref{h}, we have 
\be
h(r,x;a)
&\simeq&
\left(1-\alpha^{2}\,r\right)
-\frac{a^2\,\alpha^{2}}{4\,r}
\left[
3\,\log (\alpha^2 \, r)+x^2\,(1-\,\alpha^2\, r)
\right]
\ .
\label{h1}
\ee 
The condition~\eqref{h=0} then admits two separate solutions, namely
\begin{subequations}
\be
r_{\rm H}^{(1)} 
&\simeq&
\alpha^{-2}
\label{rtrap}
\ee
and
\be
r_{\rm H}^{(2)}(x)
&\simeq&
\frac{a^2 \, \alpha ^2 \, (9 - 2 \, x^2)}{3\,a^2 \, \alpha^4 - 8}
\sim
a^2 \, \alpha^2\, 
\frac{2 \, x^2-9}{8}
\ ,
\label{rtrap0} 
\ee
\end{subequations}
the latter of which is clearly negative for $\alpha^2 \,a \ll 1$ (since $0\le x^2\le 1$). 
Therefore, we expect there exists a horizon, whose location $r_{\rm H}^{(1)} \simeq\rh$
is given exactly by the original (spherically symmetric) solution~\eqref{rha0}
for the unperturbed space-time.
This expectation will have to be confirmed from
the study of expansions of null geodesics on $r_{\rm H}^{(1)} \simeq\rh$,
but we should also add that this calculation does not imply uniqueness and more marginally
(outer) trapped surfaces could in principle develop.
\subsection{Marginally trapped surfaces}
\label{Sec-MOTS-1}
Let us denote by $\Sigma$ the surface defined by $r = r_{\rm H}^{(1)} \simeq \alpha ^{-2}$.
In the limit of small spheroidal deformation ({\em i.e.\/}, for $\alpha^2 \,a \ll 1$),
one can easily obtain the tangent vectors to the outgoing and ingoing null geodesics on $\Sigma$,
from the conditions that they are light-like, $\bm{\ell} ^2 = \bm{n}^2 = 0$, and the normalization
$\bm{\ell} \cdot \bm{n} = -1$.
Specifically, for a metric of the form~\eqref{1}, we have
\begin{subequations}
\be
\bm{\ell}
&=&
\frac{1}{2} \, \bm{\partial}_t
+ \frac{h(r,\theta ;a)}{2}\,\sqrt{\frac{r^2+a^2}{r^2+a^2\,\cos^2\theta}} \,\bm{\partial} _r
\ ,
\label{gn3a}
\\
\bm{n}
&=&
\frac{1}{h(r,\theta;a)} \, \bm{\partial}_t 
-\sqrt{\frac{r^2+a^2}{r^2+a^2\,\cos^2\theta}} \, \bm{\partial} _r
\ ,
\label{gn3b}
\ee
\end{subequations}
since the components $\ell^{\theta}$ and $n^\theta$ (as well as $\ell^{\phi}$ and $n^\phi$)
can be set to zero without lost of generality on $\Sigma$ (for the details, see~\ref{App-G1}).
\par
Now, let $\lambda$ be the affine parameter of the null geodesics that should emerge from $\Sigma$
(to wit, $\bm{\ell} \equiv \bm{\partial}_\lambda$).
One can recast the metric~\eqref{1} in outgoing Eddington-Finkelstein-like coordinates, namely
\be 
\d s^2
&=&
- \frac{h(r,\theta ;a)}{4} \, \d \lambda ^2
+ \sqrt{\frac{r^2+a^2\,\cos^2\theta}{r^2+a^2}} \, \d r \, \d \lambda
\\
\notag
&&
+\left(r^{2}+a^{2}\cos{}^{2}\theta\right) \d\theta^{2}
+\left(r^{2}+a^{2}\right)\sin^{2}\theta\,\d\phi^{2}
\ ,
\ee
and the only non-vanishing components of the induced metric on $\Sigma$ are simply given by
$q_{22} = r^{2}+a^{2}\cos{}^{2}\theta $ and $q_{33} = \left(r^{2}+a^{2}\right)\sin^{2}\theta$.
This implies that the determinant $q = q_{22} \, q_{33}$ is a regular function of $r$ and $\theta$.
\par
From Eq.~\eqref{ThetaA} one can then conclude that
\be
\Theta _{\bm n}
=
\ell^\mu\,\frac{\partial _\mu \sqrt{q}}{\sqrt{q}}
=
\ell ^r \,\frac{\partial _r \sqrt{q}}{\sqrt{q}}
= 
- \frac{r \left(2\,r^2+a^2+a^2\,\cos^2\theta \right)}
{\sqrt{(r^2+a^2)(r^2+a^2\,\cos^2\theta)^3}}
\ ,
\ee
which is always negative, and analogously 
\be
\Theta_{\bm{\ell}}
=
-h(r,\theta;a)\,\frac{\Theta_{\bm{n}}}{2}
\ ,
\ee
which is positive for $r>\rh^{(1)}$ and vanishes on the spheroid $\Sigma$, thus confirming our initial
conjecture. 
\subsection{Misner-Sharp mass}
\label{ss:ms}
In the example considered in this section, we have found two results:
first of all, the location of the MOTS is given by the same value of the radial coordinate
as for the isotropic case (with $a=0$).
In particular, we have seen that 
\be
\Theta_{\bm{\ell}}
=
-h(r,\theta;a)\,\frac{\Theta_{\bm{n}}}{2}
\ ,
\ee
for all angles $\theta$;
the second result is that $h(\rh,\theta;a)=0$ exactly where the spherically symmetric
$h(\rh,\theta;a=0)=0$.
\par
Putting the two results together, we then find that
\be
2\,m(\rh,\theta;a)
=
2\,m(\rh)
=
\rh
\ ,
\ee
where $m(r)=m(r,\theta;a=0)$.
We can therefore conjecture that the relevant mass function for determining the location
of MOTS's in (slightly) spheroidal systems is given by the Misner-Sharp mass 
computed according to Eq.~\eqref{M} on the reference isotropic space-time.
This conjecture is somewhat reminiscent of the property of the original Misner-Sharp mass
that it is given by the volume integral over the flat reference space.
\section{A non-spheroidal source}
\label{Sec:non-spheroidal}
In this section, we want to consider the more complex case of a localised source of spheroidal radius
$r=r_0$, with mass $M_0$ and charge $Q$, surrounded by its static electric field,
with energy-momentum tensor $T^{(Q)}_{\mu\nu}$, and a suitable (electrically neutral) fluid.
We are again not interested in the inner structure of the central source, but only in the portion of space
for $r>r_0$, where we assume the metric is of the form given in Ref.~\cite{1}, that is Eq.~\eqref{1} with
\be
h (r, \theta) = h (r)
=
1-\frac{2\,M}{r}+\frac{Q^{2}}{r^{2}}
\ ,
\label{1a}
\ee
where $M$ is the total ADM mass.
\par
It is clear that the deformation parameter $a$ now measures deviations from the spherically symmetric
Reissner-Nordstr\"om metric. 
In this respect, it is worth stressing that such a deformation should be regarded as a simple example
of a (most likely) unstable intermediate configuration~\cite{Israel} within the framework of a dynamical
gravitational collapse.
\par
One can easily compute the corresponding energy-momentum tensor $T_{\mu\nu}$ by means of
the Einstein equations~\eqref{2}, from which one can see that it splits into two separate contributions,
respectively proportional to the charge $Q$ and the mass $M$,
\be
T^{\mu} _{\ \nu} = M \, {T^{(M)}} ^{\mu} _{\ \nu}  + Q^2 \, {T^{(Q)}} ^{\mu} _{\ \nu}
\ ,
\ee
which will be analysed separately.
For the part of the energy-momentum tensor associated to $M$, we consider an anisotropic fluid form, 
\begin{equation}
M \, {T^{(M)}}^{\mu}_{\ \nu}
=
\left(\varrho+p\right) u^{\mu}\,u_{\nu}
+p\,\delta_{\nu}^{\mu}
+\Pi_{\nu}^{\mu}
\ ,
\end{equation}
where $\varrho$ is the energy density, $p$ the radial pressure,
$\bm{u}$ the time-like 4-velocity of the fluid and $\Pi^{\mu\nu}$ the traceless pressure tensor
orthogonal to $\bm{u}$, 
\be
\Pi_{\ \mu}^{\mu} = \Pi_{\mu\nu}\,u^{\nu}
=0 
\ .
\label{21}
\ee
Since the system is static, we can take
\be
\bm{u}
=
\left(1-\frac{2\,M}{r}+\frac{Q^{2}}{r^{2}}\right)^{-1/2}
\, \bm{\partial}_t
\ .
\label{u0}
\ee
In particular, the (only) relevant component of the energy-momentum tensor, as far as our argument is concerned,
reads
\be
T^{0} _{\ 0}
&=&
-\frac{a^2\,M \left[\left(\cos^2 \theta - 3\right)
\left(r^{2}+a^{2}\,\cos^{2}\theta\right)
+ 4\left(r^2 + a^2\right) \cos ^2 \theta \right]}
{8\, \pi\,r \left(r^{2}+a^{2}\cos^{2}\theta\right)^{3}}
\nonumber
\\
&&
-
Q^2\,
\frac{2\left( r^2 + a^2\right) - r^2 - a^2 \,\cos ^2 \theta}
{8\, \pi\,\left(r^{2}+a^{2}\,\cos^{2}\theta\right)^{3}}
\nonumber
\\
&=&
-\varrho
+
Q^2\, {T^{(Q)}}^{0}_{\ 0}
\ .
\label{3}
\ee
\par
We only wish to discuss what happens for small deviation from the spherical symmetry,
and will therefore assume $a^{2} \ll Q^2 \ll r_0^2$.
As a further simplification, we also take $Q^2 \ll M^2$, so that the Reissner-Nordstr\"{o}m
space-time we deform is far from the extremal configuration and admits the two
horizons 
\be
r_\pm
=
M\pm\sqrt{M^2-Q^2}
\ ,
\label{rpm}
\ee
such that $h(r_\pm)=0$.
We shall review that the condition $h(r_\pm)=0$ can also be expressed in terms of the
Misner-Sharp mass like in Eq.~\eqref{hoop} in Section~\ref{ss:admMS}.  
\par
If we then put together all the previous assumptions, in every expansion we will be allowed
to neglect terms of order $a^2\,Q^{2}$ and higher.	
At leading order in $a^2$, we get
\begin{subequations}
\be 
\varrho
&\simeq&
\frac{a^2 \, M}{16\,\pi\,r^5} \left(5\, \cos 2 \theta - 1\right)
\\
\nonumber
\\ 
{T^{(Q)}}^{0}_{\ 0}
&\simeq&
-\frac{1}{8\, \pi\,r^4}
\ ,
\ee
\end{subequations}
and the \emph{total} energy density $\rho$, up to order $a^2$, is given by
\be 
\rho (r)
\simeq
\frac{1}{8\,\pi}
	\left[\frac{Q^{2}}{r^{4}}
+\frac{a^{2}M}{2\,r^{5}}\left(5\cos2\theta-1\right)
\right]
\ ,
\label{rho}
\ee
from which one can easily see that the electrostatic contribution is constant on spheroids
of constant $r$, whereas the contribution proportional to $M$ is not.
The electrostatic contribution falls within the treatment of the
previous sections, and we are here particularly interested in analysing
the effects of the latter.
\subsection{Marginally trapped surfaces}
\label{Sec-MOTS-2}
Again, following the discussion in Section~\ref{Sec-MOTS-1}, we denote by $\Sigma$
the surfaces defined by $r = r_{\pm}$.
One then finds that the tangent vector to the outgoing null geodesics on $\Sigma$
can be written as
\be
\bm{\ell}
=
\frac{1}{2} \, \bm{\partial}_t
+ \frac{h(r)}{2}\,\sqrt{\frac{r^2+a^2}{r^2+a^2\,\cos^2\theta}} \,\bm{\partial}_r
\ ,
\label{gn3a}
\ee
although setting $\ell^{\theta}=0$ on $\Sigma$ is now more subtle than it was
for the case in Section~\ref{Sec:spheroidal} and cannot be realised for
null geodesics in general (for the details, see~\ref{App-G2}).
By repeating the same steps as in Section~\ref{Sec-MOTS-1}, we thus obtain
\be
\Theta_{\bm{\ell}}
=
-h(r)\,\frac{r \, \left(2\,r^2+a^2+a^2\cos^2\theta \right)}{\sqrt{(r^2+a^2)(r^2+a^2\,\cos^2\theta)^3}}
\ ,
\ee
which again vanishes on $\Sigma$. 
\subsection{Misner-Sharp and ADM mass}
\label{ss:admMS}
We should finally recall that the Misner-Sharp mass for the isotropic Reissner-Nordstr\"om
space-time is given by (see~\ref{app:A}) 
\be
m(r)
\simeq
M
-\frac{Q^{2}}{2\,r}
\ ,
\label{mRN}
\ee
and the condition $h(r_\pm)=0$ that yields the horizons~\eqref{rpm} can indeed be written in
the form of Eq.~\eqref{hoop}, that is $2\,m(r_\pm)=r_\pm$.
This means that the results of the above analysis for the metric~\eqref{1a} do not really differ
from those for the de~Sitter space-time in Section~\ref{s:desitter}, and the isotropic
Misner-Sharp mass remains a precious indicator of the location of horizons.
In this perspective, it actually appears just like an accident that the asymptotic ADM mass
computed for the isotropic reference space-time (obtained by setting $a^2=0$) also
determines the location of the horizons. 
\par
The conjecture that the isotropic Misner-Sharp mass determines the location of slightly
spheroidal horizons nonetheless remains somewhat surprising, if one considers that the above
isotropic Misner-Sharp mass $m=m(r)$ does not coincide with the Misner-Sharp mass
adapted to the surfaces of symmetry of the spheroidal geometry.
The latter is also computed in~\ref{app:A}, where we show that it coincides with
the Hawking quasi-local mass for the system.
\section{Conclusions and outlook}
We have considered small spheroidal deformations of static isotropic systems
and studied how the MOTS is correspondingly deformed. 
Our main motivation for this investigation is to generalise the HQM~\cite{fuzzyh}
beyond the spherical symmetry, for which we need a way to locate the horizon
from quantities solely determined by the quantum state of the source.
By analysing a purely spheroidal system in section~\ref{Sec:spheroidal}, we
conjectured that such a quantity is given by the isotropic Misner-Sharp mass,
obtained by simply taking the deformation parameter to zero.
More details about the quantum description are given in Ref.~\cite{letter},
where the formalism is described for the spheroidal de~Sitter space of
Section~\ref{Sec:spheroidal}.
\par
One can understand our results about the MOTS in Section~\ref{Sec:spheroidal}
by noticing that the coordinates are deformed so as to respect the symmetry, and similar
conclusions were in fact reached in Refs.~\cite{r1-2}.
However, in section~\ref{Sec:non-spheroidal}, we have considered a spheroidal deformation of the
Reissner-Nordstr\"om metric for which a similar result is found for the location of the MOTS,
although the energy-momentum tensor of the corresponding source also contains a non-spheroidal
component.
This suggest that the general situation is very rich.
\par
It is finally important to remark again that, despite the classical instability~\cite{Israel}
of the last example, it is still possible that such a configuration appears
as an intermediate step during the collapse that leads to the formation of a
black hole.
In any case, one should not {\em a priori\/} exclude that it has a
non-vanishing probability to be realised at the quantum level
(described by the HQM).
In fact, we recall that the quantum description is the main reason of our interest
in this kind of (small) spheroidal deformations.
\section*{Acknowledgments}
R.C.~and A.G.~are partially supported by the INFN grant FLAG and
their work has been carried out in the framework of GNFM and INdAM
and the COST action {\em Cantata\/}. 
R.R.~is supported by a IRSIP grant of the HEC.
\appendix
\section{Null geodesics for slightly spheroidal de~Sitter}
\label{App-G1}
We here study in some details the null geodesics for the metric in Section~\ref{s:desitter}.
Since the $g_{xx}$ component of the metric~\eqref{gx} is not well defined at $x=1$
($\theta=0$), it will be more convenient to work with the metric in the form given originally in Eq.~\eqref{1}.
In particular, the function $h$ in Eq.~\eqref{h1} reads 
\be
h(r,\theta;a)
&\simeq&
\left(1-\alpha^{2}\,r\right)
-\frac{a^2\,\alpha^{2}}{4\,r}
\left[3\ln (\alpha^2 \, r)+\cos^{2}\theta\,(1-\,\alpha^2\, r)
\right]
\ .
\quad
\label{h3}
\ee
\par
The Lagrangian $2\,\mathcal{L}=g_{\mu\nu}\,\dot{x}^{\mu}\,\dot{x}^{\nu}$
for a point particle moving on this space-time can be written as
\be
2\,\mathcal{L}
&=&
-h(r,\theta;a)\,\dot{t}^{2}
+
\frac{r^{2}+a^{2}\cos^{2}\theta}{r^{2}+a^{2}}\,
\frac{\dot{r}^{2}}{h(r,\theta;a)}
\nonumber
\\
&&
+\left(r^{2}+a^{2}\cos{}^{2}\theta\right) \dot{\theta}^{2}
+\left(r^{2}+a^{2}\right)\sin^{2}\theta\,\dot{\phi}^{2}
\ ,
\label{h4}
\ee
where a dot represents the derivative with respect to the parameter $\lambda$
along the trajectories.
Since $t$ and $\phi$ are cyclic variables, one has the conserved conjugate momenta
\begin{subequations}
\be
p_{t}
&=&
-h(r,\theta;a)\,\dot{t}
=
-E
\ ,
\label{h4a}
\\
p_{\phi}
&=&
\left(r^{2}+a^{2}\right)\sin^{2}\theta\,\dot{\phi}
=
J
\ ,
\label{h5}
\ee
\end{subequations}
with constant $E$ and $J$, and one can always set $\dot\phi\sim J=0$.
\par
For purely radial geodesics to exist about $\Sigma$, the equation of motion for
$\theta=\theta(\lambda)$ with $J=0$, which reads 
\be
&
2 \, h^2 
\left[
\left(r^2 + a^2 \, \cos^2 \theta\right)\ddot \theta
+2 \, r \, \dot r \, \dot \theta
-
a^2 \cos \theta \, \sin \theta \, \dot \theta^2
\right]
&
\nonumber
\\
&= 
\left[ 2 \, a^2 \, h\, \cos \theta \, \sin \theta  
+ \left(r^2 + a^2 \, \cos^2 \theta\right) \partial _\theta h \right]
\left(h \, \dot \theta ^2- \frac{E^2}{r^2 + a^2 \, \cos^2 \theta} \right)
+E^2 \, \partial _\theta h
\ ,
\quad
&
\label{ddtheta}
\ee
must admit solutions with $\theta(\lambda)=\theta_0$ and (at least locally)
constant.
We then notice that
\be
\partial_\theta h
\simeq
\frac{a^2\,\alpha^{2}}{2\,r}\,(1-\,\alpha^2\, r)\,\sin\theta\,\cos\theta
\ ,
\label{dh}
\ee
so that Eq.~\eqref{ddtheta} is trivially satisfied for $\theta=0$ or $\theta=\pi$
(corresponding to a motion along the axis of symmetry) and for $\theta=\pi/2$
(motion on the equatorial plane).
Moreover, for a general value of the angular coordinate $\theta$, Eq.~\eqref{dh} ensures that
$\partial_\theta h=h=0$ on $\Sigma$ (since this surface is defined by $\alpha^2\,r=1$)
and Eq.~\eqref{ddtheta} again reduces to an identity on $\Sigma$.
This shows that radial null geodesics exist everywhere in a neighbourhood of $\Sigma$
and can be straightforwardly used to determine that $\Sigma$ is indeed a MOTS.
\section{Null geodesics for the non-spheroidal source}
\label{App-G2}
In this section we will study the radial null geodesics for the metric in Section~\ref{Sec:non-spheroidal},
which can be obtained from the Euler-Lagrange equations for the Lagrangian 
\be
2\,\mathcal{L}
&=&
- h(r) \,\dot{t}^{2}
+
\frac{r^{2}+a^{2}\cos^{2}\theta}{r^{2}+a^{2}}
\frac{\dot{r}^{2}}{h(r)}
\nonumber
\\
&&
+\left(r^{2}+a^{2}\cos^{2}\theta\right) \dot{\theta}^{2}
+\left(r^{2}+a^{2}\right)\sin^2\theta\,\dot{\phi}^{2}
\ ,
\label{34}
\ee
with $h=h(r)$ given in Eq.~\eqref{1a}.
Since $t$ and $\phi$ are cyclic variables for the Lagrangian~\eqref{34},
we still have the conserved momenta
\begin{subequations}
\be
p_{t}
&=&
-h(r)\,\dot{t}
=
-E
\ ,
\label{39a}
\\
p_{\phi}
&=&
\left(r^{2}+a^{2}\right)\sin^{2}\theta\,\dot{\phi}
=
J
\ ,
\label{39}
\ee
\end{subequations}
where $E$ and $J$ are constants.
In particular, we can always set $\dot\phi\sim J=0$.
\par
Whether the space-time at hand admits radial geodesics can be
determined from the dynamical equation for $\theta=\theta(\lambda)$
with $J=0$, that is
\be
\left(r^{2}+a^{2}\cos^{2}\theta\right)
\ddot\theta
=
-2\,r\,\dot {r}\,\dot\theta
-\frac{a^{2}\sin\theta\cos\theta}{r^{2}+a^{2}}
\frac{\dot{r}^{2}}{h(r)}
+a^{2}\sin\theta\cos\theta\,\dot{\theta}^{2}
\ .
\label{35}
\ee
In particular, upon setting $\dot\theta=0$, we obtain 
\be
\ddot\theta
=
-\frac{a^{2}\sin\theta\cos\theta}
{(r^{2}+a^{2})\left(r^{2}+a^{2}\cos^{2}\theta\right)}
\frac{\dot{r}^{2}}{h(r)}
\ ,
\label{35h}
\ee
which yields $\ddot\theta=0$ on the equatorial plane (at $\theta={\pi}/{2}$)
and along the axis of symmetry (at $\theta=0$ or $\theta=\pi$). 
For a general angular coordinate $\theta$, however, $\ddot\theta$ diverges for $r \to r _\pm$,
unless $\dot r^2/h\simeq 0$ on $\Sigma$.
Of course, the latter condition must be satisfied by outgoing null geodesics if $\Sigma$ is indeed
a horizon.
We moreover note that $r=r(\lambda)$, with $\dot\theta=\dot\phi=0$, can be obtained
from $\mathcal{L}=0$ and reads
\be
\dot{r}^{2}
=
\frac{r^{2}+a^{2}}{r^{2}+a^{2}\cos^{2}\theta}\,
h^2(r) \,\dot{t}^{2}
=
\frac{r^{2}+a^{2}}{r^{2}+a^{2}\cos^{2}\theta}\,
E^{2}
\ ,
\ee
so that $\dot r^2\propto h^2(r)=0$ on $\Sigma$ implies that $E=0$ and $\dot t$ is arbitrary 
for $r=r_\pm$.
\par
Let us compare with the behaviour of null geodesics in the Schwarzschild space-time,
for which the Lagrangian is given by
\be 
2 \, \mathcal{L}
=
- f(r) \, \dot{t}^2 + \frac{\dot{r}^2}{f(r)}
+ r^{2} \left( \dot{\theta}^{2} + \sin^2\theta\,\dot{\phi}^{2}\right)
\ ,
\label{Lschw}
\ee
with $ f(r) = 1 - {2 \, M}/{r}$.
One immediately finds the conserved quantities
\begin{subequations}
\be 
f(r) \, \dot{t}
&=&
E
\label{Eschw}
\\
r^2 \, \sin^2 \theta \, \dot{\phi}
&=&
J
\ , 
\ee
\end{subequations}
and the equation for $\theta=\theta(\lambda)$ then reads~\footnote{We do not employ the freedom to
rotate the reference frame so that the geodesic motion occurs on the equatorial plane
$\theta=\pi/2$ precisely for the purpose of comparing with the spheroidal case.}
\be 
\ddot{\theta}
=
- \frac{2 \, \dot{\theta} \, \dot{r}}{r}
+ \frac{J^2\,\cos \theta}{r^2\,\sin\theta}
\ .
\label{ddthetaSchw}
\ee
This equation clearly admits as solution $\ddot{\theta} = \dot\theta =0$ for $\dot\phi\sim J=0$.
Hence, $\mathcal{L}=0$ yields the radial equation of motion
\be 
\dot{r} = \pm E
\ .
\ee
From Eq.~\eqref{Eschw}, it is easy to see that for $f = 0$ one can set  $E = \dot r= 0$
and therefore $r = 2 \, M$ and constant is a solution.
We conclude that $\dot{t}$ is arbitrary for null geodesics trapped on the surface $\Sigma$, as above.
\par
Finally, we remark that radial geodesics with $E\not=0$ do not show any pathology in Schwarzschild,
since $\ddot\theta$ in Eq.~\eqref{ddthetaSchw} remains finite on $\Sigma$.
The same geodesics should satisfy Eq.~\eqref{35h} in the metric of Section~\ref{Sec:non-spheroidal},
namely
\be
\ddot\theta
=
-\frac{a^{2}\sin\theta\cos\theta}
{\left(r^{2}+a^{2}\cos^{2}\theta\right)^2}
\frac{E^{2}}{h(r)}
\ ,
\ee
which instead diverges on $\Sigma$ (with the exception of the equatorial plane and the symmetry axis).
\section{Mass functions}
\label{app:A}
We here compute the Hawking mass for the metric of Section~\ref{Sec:non-spheroidal},
and show that it coincides with a Misner-Sharp mass adapted to the spheroidal symmetry.
\subsection{Hawking(-Hayward) mass}
The general Hawking-Hayward mass~\cite{3} is defined as the surface integral
\begin{equation}
\bar M
=
\frac{\mathcal{A}^{1/2}}{32\,\pi^{3/2}}
\int\limits_{\Sigma}
\mu
\left[{ \mathcal{R}}
+\Theta_{+}\Theta_{-}
-\frac{\sigma_{\alpha\beta}^{+}\,\sigma_{-}^{\alpha\beta}}{2}
-2\,\omega_{\alpha}\,\omega^{\alpha}\right]
\qquad
\label{24}
\end{equation}
where $\mathcal{R}$ denotes the induced Ricci scalar on the 2-surface $\Sigma$,
$\Theta_{(\pm)}$ and $\sigma_{\alpha\beta}^{(\pm)}$ denote the expansion scalars and
shear tensors of a pair of outgoing and ingoing null geodesic congruences from
the surface $\Sigma$, respectively, $\omega_{\alpha}$ is the projection onto $\Sigma$ of the
commutator of the null normal vectors to $\Sigma$, $\mu$ is the volume 2-form on $\Sigma$
and $\mathcal{A}$ the area of $\Sigma$.
For the metric~\eqref{1a}, one immediately finds that $\omega_{\alpha}$ is of order
$a^2$ and, since we are considering all expressions only up to order $a^2$, 
the last term can be dropped.
The Hawking-Hayward mass then reduces to the Hawking mass, which we are now
going to determine.
\par
According to the general contracted Gauss equation~\cite{2}
\begin{equation}
\mathcal{R}
+\Theta_{+}\,\Theta_{-}
-\frac{1}{2}\,\sigma_{\alpha\beta}^{+}\,\sigma_{-}^{\alpha\beta}
=
h^{\alpha\gamma}\,h^{\beta\delta}\,R_{\alpha\beta\gamma\delta}
\ ,
\label{24a}
\end{equation}
where $h^{\alpha\gamma}$ is the induced metric on the 2-surface $\Sigma$ and
$R_{\alpha\beta\gamma\delta}$ is the Riemann tensor.
The tensor $h^{\alpha\gamma}$\ can be written as
\begin{equation}
h^{\alpha\gamma}
=
g^{\alpha\gamma}+l^{\alpha}\,n^{\gamma}+l^{\gamma}\,n^{\alpha},
\label{25}
\end{equation}
where $g^{\alpha\gamma}$ is the inverse of metric tensor $g_{\alpha\gamma}$,
$l^{\alpha}$ and $n^{\alpha}$ are null vectors.
On expanding in powers of $a^{2}$,
\begin{subequations}
\be
h^{\alpha\gamma}
&\simeq&
h_{0}^{\alpha\gamma}+a^{2}h_{1}^{\alpha\gamma}\ ,
\label{26}
\\
R_{\alpha\beta\gamma\delta}
&\simeq&
R_{\alpha\beta\gamma\delta(0)}+a^{2}R_{\alpha\beta\gamma\delta(1)}
\ ,
\label{27}
\ee
\end{subequations}
we obtain
\be
h^{\alpha\gamma}\,h^{\beta\delta}\,R_{\alpha\beta\gamma\delta}
&\simeq&
h_{0}^{\alpha\gamma}\,h_{0}^{\beta\delta}\,R_{\alpha\beta\gamma\delta(0)}
+a^{2}\,h_{0}^{\alpha\gamma}\,h_{0}^{\beta\delta}R_{\alpha\beta\gamma\delta(1)}
\nonumber
\\
&&
+a^{2}
\left(h_{1}^{\alpha\gamma}\,h_{0}^{\beta\delta}+h_{0}^{\alpha\gamma}\,h_{1}^{\beta\delta}\right)
R_{\alpha\beta\gamma\delta(0)}
\ .
\label{28}
\ee
The components of $h^{\alpha\gamma}$ are determined from Eqs.~(\ref{25})
and (\ref{26}), which gives
\be
h_{0}^{\alpha\gamma}+a^{2}h_{1}^{\alpha\gamma}
&\simeq&
g_{0}^{\alpha\gamma}+a^{2}\,g_{1}^{\alpha\gamma}
+(l_{0}^{\alpha}+a^{2}\,l_{1}^{\alpha})
(n_{0}^{\gamma}+a^{2}n_{1}^{\gamma})
\nonumber
\\
&&
+(l_{0}^{\gamma}+a^{2}l_{1}^{\gamma})(n_{0}^{\alpha}+a^{2}n_{1}^{\alpha})
\ ,
\ee
where $g_{0}^{\alpha\gamma}$ and $g_{1}^{\alpha\gamma}$ are zeroth and first
order terms of the metric tensor respectively, and similarly for $l_{0}^{\alpha}$,
$l_{1}^{\alpha}$ and $n_{0}^{\alpha}$, $n_{1}^{\alpha}$.
For the unperturbed Reissner-Nordstr\"om space-time $l_{0}^{\alpha}=n_{0}^{\alpha}=0$, for $\alpha=2$ and $3$, 
so that, up to order $a^2$, we have
\begin{subequations}
\be
h_{0}^{22}+a^{2}\,h_{1}^{22}
&=&
g_{0}^{22}+a^{2}g_{1}^{22}+2\,a^{4}\,l_{1}^{2}\,n_{1}^{2}
\nonumber
\\
&\simeq&
g_{0}^{22}+a^{2}\,g_{1}^{22}
\ ,
\\
h_{0}^{33}+a^{2}\,h_{1}^{33}
&=&
g_{0}^{33}+a^{2}g_{1}^{33}+2\,a^{4}\,l_{1}^{3}\,n_{1}^{3}
\nonumber
\\
&\simeq&
g_{0}^{33}+a^{2}\,g_{1}^{33}
\ .
\ee
\end{subequations}
In particular, for the metric~(\ref{1a}), we find
\begin{equation}
h^{22}
\simeq
\frac{1}{r^{2}}-
a^2\,\frac{\cos^{2}\theta}{r^{4}}
\ .
\label{28a}
\end{equation}
Similarly,
\begin{equation}
h^{33}
\simeq
\frac{1}{r^{2}\sin^{2}\theta}
-\frac{a^2}{r^{4}\sin^{2}\theta}
\ .
\label{28b}
\end{equation}
Eq.~(\ref{28}) now reduces to
\be
h^{\alpha\gamma}\,h^{\beta\delta}\,R_{\alpha\beta\gamma\delta}
&\simeq&
2\left[h_{0}^{22}\,h_{0}^{33}\,R_{2323(0)}
+a^{2}\,h_{0}^{22}\,h_{0}^{33}\,R_{2323(1)}
\right.
\nonumber
\\
&&
\left.
+a^{2}\left(h_{1}^{22}\,h_{0}^{33}+h_{0}^{22}\,h_{1}^{33}\right)
R_{2323(0)}\right]
\ ,
\ee
where 
\be
R_{2323}
&=&
\frac{\sin^{2}\theta(r^{2}+a^{2})(2Mr-q^{2})}{r^{2}+a^{2}\cos^{2}\theta}
\nonumber
\\
&\simeq&
\left(2Mr-q^{2}\right)\sin^{2}\theta
+\frac{2a^{2}M\sin^{4}\theta}{r}
\ .
\label{28c}
\ee
By using equations (\ref{28a}-\ref{28c}), the contracted Gauss equation~(\ref{24a})
takes the form
\be
h^{\alpha\gamma}\,h^{\beta\delta}\,R_{\alpha\beta\gamma\delta}
&=&
\mathcal{R}
+\theta_{+}\,\theta_{-}
-\frac{1}{2}\,\sigma_{\alpha\beta}^{+}\,\sigma_{-}^{\alpha\beta}
\nonumber
\\
&=&
\frac{2\left(2Mr-q^{2}\right)}{r^{4}}
-\frac{8a^{2}M\cos^{2}\theta}{r^{5}}
\ .
\label{29}
\ee
Since $h_{22}=g_{22}=$($r^{2}+a^{2}\cos^{2}\theta)$ and
$h_{33}=g_{33}=(r^{2}+a^{2})\sin^{2}\theta$, the volume 2-form 
\be
\mu
&=&
\sqrt{\det(h_{\alpha\beta})}
\,d\theta\, d\phi
\nonumber
\\
&\simeq&
\left(r^{2}
+a^2\,\frac{3+\cos2\theta}{4}
\right)
\sin\theta\,d\theta\, d\phi
\ .
\label{29b}
\ee
The area of $\Sigma$ is then given by
\begin{equation}
\mathcal{A}
=
\int\limits_{\Sigma}
\mu
\simeq
4\,\pi\left(r^{2} + \frac{2}{3}\,a^2\right)
\ .
\label{30}
\end{equation}
The Hawking mass is finally obtained from Eq.~(\ref{24}) with Eqs.~\eqref{29} and \eqref{30},
which yields
\be
M_{\rm H}(r)
\simeq
M
\left(1+\frac{a^{2}}{3\,r^{2}}\right)
-\frac{q^{2}}{2\,r}
\ .
\label{31}
\ee
We are next going to recover this result in a different way.
\subsection{Adapted Misner-Sharp mass?}
The Misner-Sharp mass~\eqref{M} is properly defined only for spherically symmetric
space-times.
One could generalise it by integrating the matter density on the spatial volume
inside surfaces of symmetry, which are given by spheroids in the present case.
In other words, we replace Eq.~\eqref{M} with
\be
m(r)
=
M_0
+
{2\pi}
\int\limits_{0}^{\pi}
\displaystyle\int\limits_{r_0}^{r}
\sqrt{\gamma(\bar r,\theta)}\,\rho(\bar r,\theta)\,
\d\bar r\,\d\theta
\ ,
\label{23a}
\ee
where we recall that $r=r_0$ is the coordinate of the inner core, and
$\gamma=(r^{2}+a^{2}\cos^{2}\theta)^2\sin^2\theta$ is the determinant
of the flat 3-metric in spheroidal coordinates,
\be
\gamma_{ij}\,\d x^i\,\d x^j
=
\frac{r^{2}+a^{2}\cos{}^{2}\theta}{r^{2}+a^{2}}\,\d r^{2}
+\left(r^{2}+a^{2}\cos^{2}\theta\right) \d\theta^{2}
+\left(r^{2}+a^{2}\right)\sin^{2}\theta\,\d\phi^{2}
\ .
\ \ \ \
\ee
Eq.~\eqref{23a} then yields
\be
m(r)
\simeq
M_0
-\frac{Q^{2}}{2}\left(\frac{1}{r}-\frac{1}{r_0}\right)
+\frac{a^{2}\,M}{3}\left(\frac{1}{r^{2}}-\frac{1}{r_0^{2}}\right)
\ .
\quad
\label{mh}
\ee
For $r\to\infty$, the above expression should equal the total ADM mass $M$,
that is
\be
M
\simeq
M_0
-\frac{M\,a^{2}}{3\,r_0^2}
+
\frac{Q^{2}}{2\,r_0}
\ .
\ee
This allows us to express $r_0$ and $M_0$ so that Eq.~\eqref{mh} becomes
\be
m(r)
\simeq
M
\left(1+\frac{a^{2}}{3\,r^2}\right)
-\frac{Q^{2}}{2\,r}
=
M_{\rm H}(r)
\ ,
\label{mH}
\ee
from which the isotropic Misner-Sharp mass~\eqref{mRN} is obtained by taking $a^2\to 0$.
\par
This calculations therefore shows that, at least for spheroidal space-times like~\eqref{1a},
one can expect the Hawking mass function evaluated on surfaces of symmetry equals
the adapted Misner-Sharp function evaluated inside volumes bounded by the same surfaces
of symmetry.

\end{document}